\documentclass{raa}
\usepackage{graphicx,times}
\usepackage{natbib}
\usepackage{amssymb,amsmath}
\bibpunct{(}{)}{;}{a}{}{,}

\usepackage[a4paper=true,dvipdfm=true,pagebackref=true]{hyperref}
\hypersetup{pdftitle = The title of my PDF, pdfauthor = My name, pdfsubject= The subject, pdfkeywords = keyword1 keyword2 keyword3}
\hypersetup{colorlinks = true, linkcolor = green, anchorcolor = red, citecolor = blue, filecolor = red, pagecolor = red, urlcolor = red}

\newcommand{\FeH}{$\mbox{[Fe/H]}$} 
\newcommand{\AB}[2]{$\mbox{[#1/#2]}$} 

\newcommand{\FeHeq}[1]{$\mbox{[Fe/H]}={#1}$}    
\newcommand{\FeHsim}[1]{$\mbox{[Fe/H]}\sim{#1}$}  
\newcommand{\ABeq}[3]{$\mbox{[#1/#2]}={#3}$}    
\newcommand{\ABlt}[3]{$\mbox{[#1/#2]}<{#3}$}    
\newcommand{\ABgt}[3]{$\mbox{[#1/#2]}>{#3}$}    
\newcommand{\ABge}[3]{$\mbox{[#1/#2]}\ge{#3}$}   
\newcommand{\ABsim}[3]{$\mbox{[#1/#2]}\sim {#3}$} 
\newcommand{\ABlesssim}[3]{$\mbox{[#1/#2]}\lesssim {#3}$} 
\newcommand{\tefft}{$T_{\mbox{\scriptsize eff}}$} 
\newcommand{\hn}[1]{\textbf{\textit{#1}}}
\newcommand{\JrII}{LAMOST~J110901.22$+$075441.8}

\newcommand{\logg}{\ensuremath{\log g}}

\newcommand{\mlp}{\ensuremath{\alpha_{\mathrm{MLT}}}}

\newcommand{\Tefft}{\ensuremath{T_{\mathrm{eff}}}}

\begin{document}

\title{Discovery of a strongly $r-$process enhanced extremely metal-poor star {\JrII}
}

\volnopage{ {\bf 2015} Vol.\ {\bf X} No. {\bf XX}, 000--000}
\setcounter{page}{1}

\author{Haining Li\inst{1}, Wako Aoki\inst{2,3}, Satoshi Honda\inst{4}, Gang Zhao\inst{1},
   Norbert Christlieb\inst{5}, Takuma Suda\inst{6}}


\institute{ Key Lab of Optical Astronomy, National Astronomical Observatories,
Chinese Academy of Sciences, A20 Datun Road, Chaoyang, Beijing 100012, China; {\it lhn@nao.cas.cn}\\
\and
     National Astronomical Observatory of Japan, 2-21-1 Osawa, Mitaka, Tokyo, 181-8588, Japan\\
\and
     Department of Astronomical Science, School of Physical Sciences, 
     The Graduate University of Advanced Studies (SOKENDAI), 2-21-1 Osawa, Mitaka, Tokyo 181-8588, Japan\\
\and
     University of Hyogo, 407-2, Nishigaichi, Sayo-cho, Sayo, Hyogo, 679-5313, Japan\\
\and
     Zentrum f{\"u}r Astronomie der Universit{\"a}t Heidelberg, Landessternwarte,
       K{\"o}nigstuhl 12, D-69117 Heidelberg, Germany\\
\and
     Research Center for the Early Universe, The University of Tokyo, Hongo 7-3-1, Bunkyo-ku, Tokyo 113-0033, Japan\\
\vs \no
   {\small Received 2015 April 02; accepted 2015 March 13}
}

\abstract{We report the discovery of an extremely metal-poor (EMP) giant, {\JrII},
which exhibits large excess of $r-$process elements with \ABsim{Eu}{Fe}{+1.16}.
The star is one of the newly discovered EMP stars identified
from LAMOST low-resolution spectroscopic survey
and the high-resolution follow-up observation with the Subaru Telescope.
Stellar parameters and elemental abundances have been determined from the Subaru spectrum.
Accurate abundances for a total of 23 elements including 11 neutron-capture elements
from Sr through Dy have been derived for {\JrII}.
The abundance pattern of {\JrII} in the range of C through Zn is in line with the ``normal''
population of EMP halo stars, except that it shows a notable underabundance in carbon.
The heavy element abundance pattern of {\JrII} is in agreement with other well studied
cool $r-$II metal-poor giants such as CS~22892$-$052 and CS~31082$-$001.
The abundances of elements in the range from Ba through Dy well match the scaled Solar $r-$process pattern.
{\JrII} provides the first detailed measurements of neutron-capture elements 
among $r-$II stars at such low metallicity with \ABlesssim{Fe}{H}{-3.4}, 
and exhibits similar behavior in the abundance ratio of \AB{Zr}{Eu} 
as well as Sr/Eu and Ba/Eu as other $r-$II stars.
}

\authorrunning{H. N. Li et al. }            
\titlerunning{Discovery of an EMP $r-$II star}  
\maketitle

%
\section{Introduction}           
\label{sect:intro}

Detailed elemental abundances of metal-poor stars in the Galactic halo
provide fundamental knowledge about the history and nature of
the nucleosynthesis in the Galaxy. Particularly, abundances of the slow ($s$)
and rapid ($r$) neutron-capture elements are of great importance
to constrain early Galactic nucleosynthesis and chemical evolution.
In the past decades, great efforts have been devoted to studies in relevant fields
based on high-resolution spectroscopic measurements of the elemental abundances of
metal-poor stars in the Galaxy.
These observations together with theoretical studies have revealed
that the $r-$process is primarily responsible for the production
of heavy elements beyond the iron group \citep{Spite&Spite1978AA,Sneden1996ApJ} in the early Galaxy.
Only at later time (higher metallicities), the onset of the $s-$process occurs,
injecting nucleosynthesis material into the interstellar medium
from long-lived low- and intermediate mass stars \citep{Burris2000ApJ}.
Detections of radioactive elements including thorium and uranium
in a few $r-$process enhanced metal-poor stars have provided an independent
measurement on the age of these oldest stars, and hence set lower limits
to the age of the Galaxy \citep{Hill2002AA,Frebel2007ApJ}.

\citet{Sneden1994ApJ} found the first $r-$process enhanced extremely metal-poor (EMP) giant
CS~22892$-$052, with a \ABsim{Eu}{Fe}{+1.6}
\footnote{$[A/B]=\log(N_A/N_B)_{\star}-\log(N_A/N_B)_{\odot}$,
where $N_A$ and $N_B$ are the number densities of elements A and B respectively,
and $\star$ and $\odot$ refer to the star and the Sun respectively}
and an abundance pattern from Ba through Dy
which is similar to a scaled Solar System $r-$process (SSr) distribution.
A few rare stars have subsequently been found to exhibit extreme enhancements in $r-$process elements,
suggesting that the observed abundances are dominated by the effect
from a single or very few nucleosynthesis events.
Based on the definition by \citet{Beers&Christlieb2005ARAA}, such stars with
\ABgt{Eu}{Fe}{+1} and \ABlt{Ba}{Eu}{0} are referred to $r-$II stars.
Although the astrophysical site of the $r-$process is not yet clear,
the $r-$process is believed to be connected to explosive conditions
of massive-star core-collapse supernovae \citep{Woosley1994ApJ},
or neutron star mergers \citep{Goriely2013PhRvL}.
Therefore, $r-$II stars are the best candidates to explore the details of the $r-$process and its site.

There are 12 $r-$II stars known to date \citep[e.g.,][]{Hill2002AA,Sneden2003ApJ,Honda2004ApJ,
Sneden2008ARAA,Hayek2009AA,Mashonkina2010AA,Aoki2010ApJL}.
Detailed abundance analysis of these $r-$II stars have found 
\hn{excellent match between the stellar and solar
$r-$process pattern in the Ba-Hf range}, which indicates a universal $r-$process,
i.e, elements were produced with the same proportions during the evolution of the Galaxy.
Such conclusion is of fundamental importance for a better understanding
of the nature of the $r-$process.
To establish the origin of the heavy elements beyond the iron group
among the oldest stars in the Galaxy, larger samples of accurate measurements
of additional elements are required.

{\JrII} was identified as a candidate of EMP stars from the first data release \citep{Luo2015arXiv}
of the low-resolution ($R\sim1800$) spectroscopic survey of LAMOST
\footnote{See http://www.lamost.org for more detailed information, and the progress of the LAMOST surveys.}
\citep[the Large sky Area Multi-Object fiber Spectroscopic Telescope,
also known as Wang-Su Reflecting Schmidt Telescope or Guoshoujing Telescope;][]
{Zhao2006ChJAA,Cui2012RAA,Zhao2012RAA,Luo2012RAA}.
The follow-up high-resolution spectroscopic observation was carried out 
with the Subaru Telescope (Li et al. 2015b),
which has confirmed its being a $r-$II EMP giant. In this paper,
we introduce the observation and measurements of parameters and abundances of {\JrII} in Section~\ref{sec:observation}; 
results and interpretations on the elemental abundance are presented in Section~\ref{sec:abundance};
patterns of the heavy elements and conclusions are described in Sections~\ref{sec:abun-pattern} and \ref{sec:conclusion}.

\section{Observations and measurements}\label{sec:observation}

\subsection{Target selection and follow-up observations}\label{subsec:target}

\begin{figure}
 \begin{center}
  \includegraphics[width=10cm]{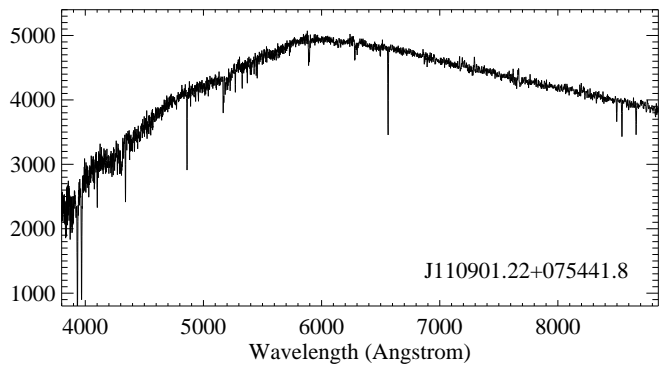}
 \end{center}
\caption{Medium-resolution spectrum of the r-II star obtained by LAMOST.}\label{fig:LAMOST-spec}
\end{figure}

{\JrII} was selected as a candidate of EMP stars from the low-resolution spectra
of the first data release of the LAMOST spectroscopic survey.
The wavelength coverage ($3700$--$9100$\,{\AA}) and resolving power ($R=1800$)
of the LAMOST spectra (Fig.~\ref{fig:LAMOST-spec}) 
allow a robust estimation of the stellar parameters including metallicities.
Methods to determine the metallicity of an object and the selection of EMP candidates
were similar to those made by \citet{Li2015ApJ}.

For 54 candidates of extremely metal-poor stars selected from the
LAMOST sample,``snapshot'' high-resolution spectra were acquired with
the resolving power $R=36,000$ and exposure times of 10$-$20 minutes
during the two-night run in May 2014 with Subaru/HDS as made by \citet{Aoki2013AJ}.
{\JrII} was observed on May 9.
Since this object is relatively bright ($r$ = 12.08),
quite high signal-to-noise ratio was achieved covering 4000$-$6800\,{\AA},
e.g., with a S/N of about 70 at 4500\,{\AA} and 110 at 5200\,{\AA}
by an exposure of 10 minutes.
Data reduction was carried out with standard procedures
using the IRAF echelle package including bias-level correction,
scattered light subtraction, flat-fielding, extraction of spectra, and
wavelength calibration using Th$-$Ar arc lines. Cosmic-ray hits were removed
by the method described in \citet{Aoki2005ApJ}.

Radial velocities of the sample were obtained using the
\texttt{IRAF} procedure \texttt{fxcor}, and a synthetic spectrum
with low-metallicity was employed as a template for cross-correlation.
The above method derives a radial velocity of $-72.3\pm0.4$\,km\,s$^{-1}$.

\subsection{Stellar parameters}\label{subsec:stellar-param}

Equivalent widths were measured by fitting Gaussian profiles to isolated
atomic absorption lines based on the line list of \citet{Aoki2013AJ}
for elements lighter than Zn, and that from \citet{Mashonkina2010AA}
for heavy elements beyond Sr.

Since there is not yet a uniform photometric system for all the LAMOST input catalogue,
we have adopted the spectroscopic method to derive the stellar parameters
of the whole sample, including {\JrII}.
By minimizing the trend of the relationship between the derived abundances
and excitation potentials of Fe\,I lines, the effective temperature {\tefft} of {\JrII}
was determined. The empirical formula derived by \citet{Frebel2013ApJ}
has been adopted to correct the usually expected systematic offsets between the
spectroscopic and photometric effective temperatures,
which resulted in a {\Tefft} of 4441\,K from the original value of 4190\,K.
The microturbulent velocity $\xi$ was also determined based on analysis of
Fe\,I lines, i.e., by forcing the iron abundances of individual lines to
exhibit no dependence on the reduced equivalent widths.
The sufficiently high quality of the Subaru spectrum allows us to detect 11 Fe\,II lines for {\JrII}.
Therefore the surface gravity {\logg} was determined by minimizing the difference
between the average abundances derived from the Fe\,I and Fe\,II lines.

The derived parameters of {\JrII} are listed in Table~\ref{tab:param-abun}.
Since the $V-K$ color is available for {\JrII} ($(V-K)_{0}=2.54$),
we have also checked the photometric temperature of this object.
Adopting the calibration by \citet{Alonso1999AAS,Alonso2001AA},
the derived temperature is 4560\,K, which is consistent with the spectroscopic temperature
within the level of the uncertainty of our measurement.

\subsection{Abundance determination}\label{subsec:abun-determination}

For the abundance analysis, the 1D plane-parallel, hydrostatic model atmospheres of
the ATLAS NEWODF grid of \citet{Castelli&Kurucz2003IAUS} were adopted,
assuming a mixing-length parameter of $\mlp=1.25$, no convective overshooting,
and local thermodynamic equilibrium. An updated version of the abundance analysis code MOOG \citep{Sneden1973ApJ} was used,
which does not treat continuous scattering as true absorption,
but as a source function which sums both absorption and scattering \citep{Sobeck2011AJ}.
When calculating [X/H] and [X/Fe] abundance ratios,
the photospheric Solar abundances of \citet{Asplund2009ARAA} were adopted .

Abundances of most elements were computed using the measured equivalent widths
of isolated atomic lines with the derived stellar parameters.
For heavy elements whose spectral lines show hyperfine splitting, such as Sr, Ba, La and Eu, 
the elemental abundances were determined using spectral synthesis taking into account
the isotopic splitting and/or hyperfine structure (HFS).
The derived abundances of {\JrII} are listed in Table~\ref{tab:param-abun},
which also includes $N$, the number of lines which have been used to determine the abundance,
together with the abundance error as described in the following text.

The uncertainties of the derived abundances mainly come from two aspects,
i.e, the uncertainties of the equivalent width measurements,
and those caused by the uncertainties of stellar parameters.
In the case of equivalent width measurement,
when $N \ge 2$ lines of individual species of an element were observed,
the dispersion around the average abundance was used to present random error;
if the elemental abundance was determined from a single line,
the statistic error of the equivalent widths was estimated
based on the classical formula of \citet[][Equation 7]{Cayrel1988IAUS}.
The abundance uncertainties associated with the uncertainties of
the stellar parameters were estimated by individually varying {\tefft} by $+$150\,K,
{\logg} by $+$0.1\,dex, and $\xi$ by $+$0.1\,km\,s$^{-1}$
in the stellar atmospheric model.
The total uncertainty of the errors have been computed as the quadratic sum of the above aspects,
and are shown in the column of $\sigma$ in Table~\ref{tab:param-abun}.

\begin{table}
\caption{Stellar Parameters and Elemental abundances of {\JrII}.}\label{tab:param-abun}
\begin{center}
\begin{tabular}{lrrrrcr}
\hline\hline
   &\multicolumn{4}{c}{{\JrII}}&&Sun\\
\cline{1-5}\cline{7-7}
{\Tefft(K)}&\multicolumn{4}{c}{4440$\pm$150}&&\\
{\logg} &\multicolumn{4}{c}{0.70$\pm$0.1}&&\\
{\FeH}  &\multicolumn{4}{c}{$-$3.41$\pm$0.1}&&\\
{$\xi$(km\,s$^{-1}$)} &\multicolumn{4}{c}{1.98$\pm$0.1}&&\\
\cline{1-5}\cline{7-7}
Ion&log\,$\epsilon$(X)&\AB{X}{Fe}&$N$&$\sigma$&&log\,$\epsilon$(X)\\
\hline
C     &    4.45& $-$0.57& 1& 0.38&&   8.43\\
Na    &    3.41&    0.58& 2& 0.23&&   6.24\\
Mg    &    4.60&    0.41& 4& 0.15&&   7.60\\
Si    &    4.58&    0.48& 1& 0.17&&   7.51\\
Ca    &    3.30&    0.31&15& 0.16&&   6.34\\
Sc    & $-$0.33& $-$0.07& 8& 0.16&&   3.15\\
Ti\,I &    1.71&    0.17&10& 0.23&&   4.95\\
Ti\,II&    1.81&    0.27&25& 0.11&&   4.95\\
Cr    &    1.83& $-$0.40& 5& 0.20&&   5.64\\
Fe\,I &    4.09&    0.00&98& 0.24&&   7.50\\
Fe\,II&    4.08& $-$0.01&11& 0.14&&   7.50\\
Co    &    1.64&    0.06& 1& 0.23&&   4.99\\
Ni    &    2.75& $-$0.06& 2& 0.16&&   6.22\\
Zn    &    1.62&    0.47& 2& 0.09&&   4.56\\
Sr    & $-$0.25&    0.29& 2& 0.19&&   2.87\\
Y     & $-$1.20&    0.00& 3& 0.12&&   2.21\\
Zr    & $-$0.54&    0.29& 2& 0.17&&   2.58\\
Ba    & $-$0.92&    0.31& 3& 0.14&&   2.18\\
La    & $-$1.60&    0.71& 2& 0.16&&   1.10\\
Ce    & $-$1.47&    0.36& 2& 0.13&&   1.58\\
Pr    & $-$1.78&    0.91& 1& 0.15&&   0.72\\
Nd    & $-$1.47&    0.52& 4& 0.16&&   1.42\\
Sm    & $-$1.67&    0.78& 2& 0.17&&   0.96\\
Eu    & $-$1.73&    1.16& 1& 0.14&&   0.52\\
Dy    & $-$1.35&    0.96& 1& 0.15&&   1.10\\
\hline
\end{tabular}
\end{center}
\end{table}

\section{Elemental Abundances and Interpretations}\label{sec:abundance}

\begin{figure}
 \begin{center}
  \includegraphics[width=16cm]{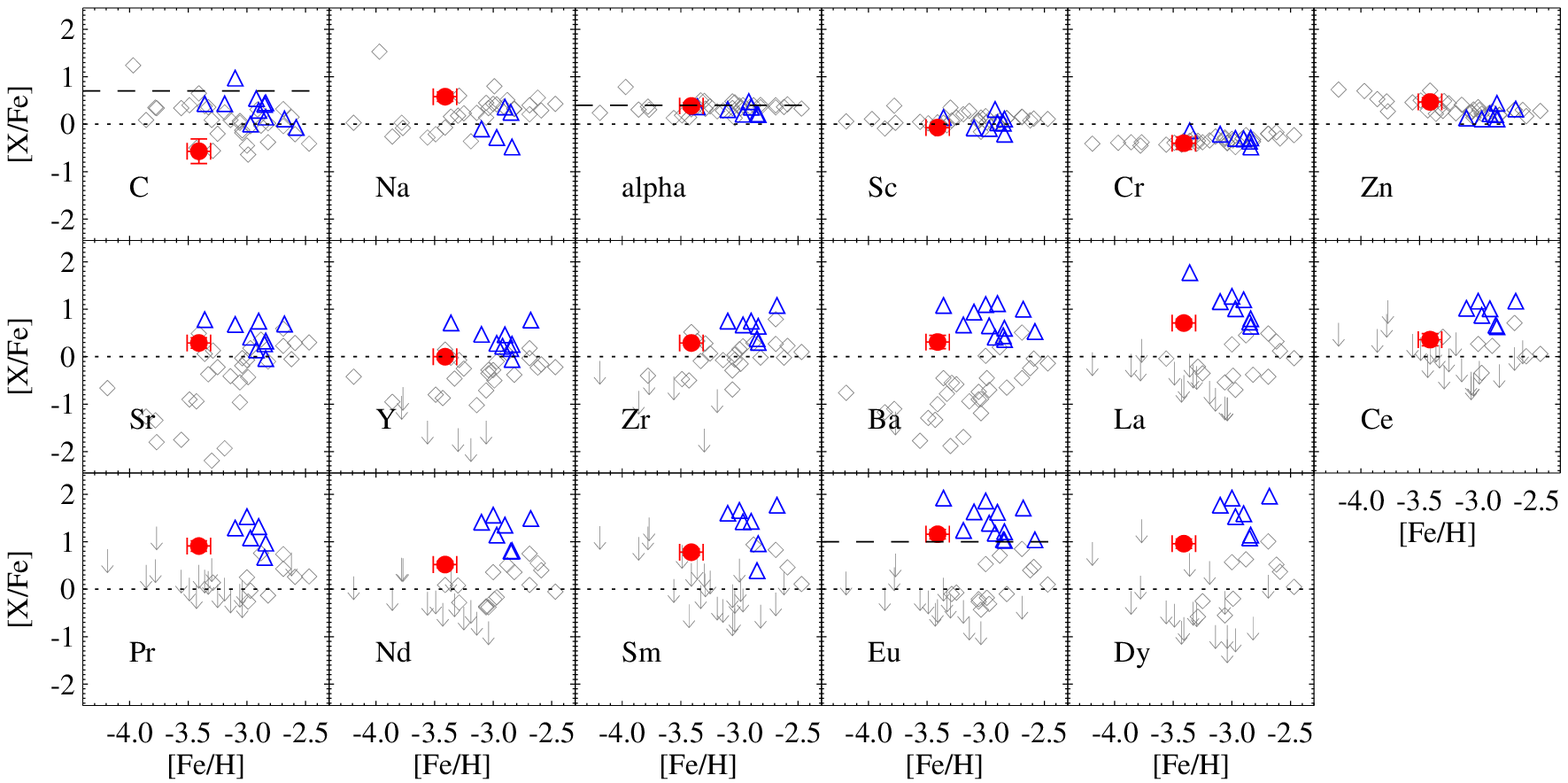}
 \end{center}
\caption{\AB{X}{Fe} vs. {\FeH} from C through Eu of {\JrII}.
Filled circles are abundances of {\JrII}.
Triangles are abundance ratios of other $r-$II stars collected from literatures
\citep{Hill2002AA,Sneden2003ApJ,Honda2004ApJ,Christlieb2004AA,Barklem2005AA,
Francois2007AA,Lai2008ApJ,Sneden2008ARAA,Hayek2009AA,Mashonkina2010AA,Aoki2010ApJL}.
Diamonds and downward arrows refer to abundances and upper-limits of metal-poor giants
from the ``First Stars'' project \citep{Cayrel2004AA,Spite2005AA,Francois2007AA}.
For C, the dashed line refers to the division between carbon-enhanced
and carbon-normal of \ABsim{C}{Fe}{+0.7} \citep{Aoki2007ApJ}.
For the $\alpha$-elements, the dashed line refers to the canonical value of \ABsim{$\alpha$}{Fe}{+0.4}
for halo stars \citep{McWilliam1997ARAA}.
For Eu, the criterion for $r-$II stars with \ABge{Eu}{Fe}{+1.0} \citep{Beers&Christlieb2005ARAA}
is also presented with a dashed line for reference.}\label{fig:abun_C2Eu}
\end{figure}

In Fig.~\ref{fig:abun_C2Eu}, abundance ratios relative to iron of {\JrII}
are shown and compared to other metal-poor $r-$II stars,
as well as the ``normal'' metal-poor giants from the ``First Stars'' project \citep{Cayrel2004AA}.
Note that there is only one cool metal-poor, main-sequence $r-$II star
discovered by \citet{Aoki2010ApJL} with \FeHeq{-3.36}, while the rest are all cool giants.

\subsection{Lithium through zinc}

{\it Lithium.} The Li\,I 6708\,{\AA} line is covered in the observed spectrum,
but we could not detect the lithium line in {\JrII}.
An upper limit of $A\mbox{(Li)}<0.35$\footnote{$A\mbox{(Li)}$ 
is defined as $A\mbox{(Li)}=\log(N\mbox{(Li)}/N\mbox{(H)})+12$.}
has been derived based on spectral synthesis.
This is not unexpected for a red giant
which is usually showing a rather low abundance of Li,
since such objects have already undergone the first dredge-up
and the surface material is mixed with internal one with depleted lithium.

{\it Carbon.} The carbon abundance of {\JrII} was derived by matching the observed
CH A-X band at 4310\,{\AA} ~(i.e., the G-band) to the synthetic spectra.
The object shows notable underabundant carbon compared to other $r-$II stars
and most of the EMP stars (Fig.~\ref{fig:abun_C2Eu}).
The derived abundance ratio \AB{C}{Fe}$=-0.57$, which is close to the ratios
of the so-called ``mixed'' EMP giants from the sample of \citet{Spite2005AA}.
Such low carbon abundance ratio is likely to be caused by mixing
which has brought processed material to the surface from deep layers.
Such cool giants should exhibit enhancement of nitrogen which is probably
converted from carbon; however, due to limited wavelength coverage
of the snapshot spectrum which does not include the NH or CN lines,
we are unable to measure the nitrogen abundance of {\JrII} with current data.

{\it Sodium.} Abundance of Na was derived based on the EW measurements of the two resonance lines.
It is noticed from Fig.~\ref{fig:abun_C2Eu} that the \ABeq{Na}{Fe}{0.58} of {\JrII} indicates relative excess 
compared to other $r-$II stars which normally show solar or slight lower \AB{Na}{Fe} ratio.
The difference may be balanced when the NLTE correction on sodium abundances are considered,
e.g., according to \citet{Cayrel2004AA}, a correction up to $-0.5$\,dex shall be added to the LTE sodium abundance.
However, we have only found \AB{Na}{Fe} from literatures for a few $r-$II stars,
and a larger sample would be needed to make any conclusive remark.

{\it The $\alpha-$elements.} For {\JrII}, abundances of four $\alpha-$elements Mg, Si, Ca, and Ti
were derived based on the EW measurements of atomic lines.
As shown in Fig.~\ref{fig:abun_C2Eu}, all the $\alpha-$elements present enhancement relative to iron,
with an average \ABeq{(Mg+Si+Ca+Ti)}{Fe}{+0.37} which agrees with the canonical value
of \ABsim{$\alpha$}{Fe}{+0.4} for halo stars \citep{McWilliam1997ARAA}.

{\it Scandium and iron$-$peak elements.} Except for V whose atomic lines are not covered
and Mn for which only one line with distorted feature could be found,
abundance of elements in the nuclear charge ranging from $Z=21$ through 28
were determined for {\JrII}.
The abundance ratio to iron of Sc, Co, and Ni are about the solar value.
Cr is deficient relative to iron and solar ratios, with \ABeq{Cr}{Fe}{-0.40},
while Zn is overabundant relative to iron with \ABeq{Zn}{Fe}{+0.47}.
These abundance ratios in {\JrII} well agree with
the general trend of EMP halo stars with similar metallicities
(with Sc, Cr, and Zn shown in Fig.~\ref{fig:abun_C2Eu} as examples).

In general, we found that the element abundance pattern of {\JrII} in the C$-$Zn range
resembles the ``normal'' pattern of halo EMP stars,
except that it shows a quite low C abundance and relatively large \AB{Na}{Fe}.
The underabundance in carbon is presumably due to the fact that
it is a more evolved star compared with EMP stars with higher temperatures.
The overabundance in sodium may be explained by the discrepancy
between LTE and NLTE Na abundances for EMP giants.

\subsection{Heavy element: strontium through dysprosium}

We determined abundances of {\JrII} for 11 heavy elements,
including three light trans-iron elements and eight elements in the region of the second $r-$process peak.
The elements representing the third peak could not be measured,
because the wavelength coverage of our spectrum did not include any strong lines for measurements.

{\it The light trans-iron elements.} Three elements were measured in the region
of the first peak with $38\le Z\le 46$, including Sr, Y, and Zr.
The abundance of Sr was measured by spectral synthesis of the two strongest resonance lines,
Sr\,II 4077 and 4215\,{\AA}. Both of the lines are affected by HFS of the odd isotope $^{87}$Sr,
and a fraction of 0.22 was adopted for this isotope for synthesis,
according to \citet{Arlandini1999ApJ} for a pure $r-$process production of strontium.
The correction of HFS and isotopic splitting has resulted in 
an increase of 0.15\,dex for the abundance of Sr, i.e., from $-0.25$ to $-0.10$.
For Y and Zr, the abundances were derived from the EWs of
three and two isolated and clear atomic lines, respectively.
Abundances of these elements in {\JrII} are quite similar to other $r-$II stars,
and are higher than the average abundance ratio of ``normal'' EMP halo stars
with similar metallicities (Fig.~\ref{fig:abun_C2Eu}).

{\it The second $r-$process peak elements.} In the region of the second peak,
abundances were determined for eight elements Ba, La, Ce, Pr, Nd, Sm, Eu, and Dy.
Four isolated barium lines could be measured for {\JrII},
including the Ba\,II 4554\,{\AA} resonance line and the three subordinate lines,
Ba\,II 5853, 6141, and 6497\,{\AA}. The three subordinate lines are almost free from HFS effects,
and derive quite similar Ba abundances with a difference of about 0.01\,dex.
However, the resonance line is strongly affected by HFS.
Even if we take the HFS splitting for spectral synthesis into account,
the Ba\,II 4554\,{\AA} line still results in a Ba abundance with 0.20\,dex
higher than that from the subordinate lines.
We suspect that this difference is mainly caused by the NLTE effect \citep{Mashonkina&Christlieb2014AA},
which is not included in our analysis for this paper.
Therefore, the adopted Ba abundance of {\JrII} was solely based on the three subordinate lines.
HFS splitting was also accounted for La \citep{Lawler2001ApJb},
which resulted in a difference of 0.3\,dex in log\,$\epsilon$(La).
Europium is another element for which HFS and isotope splitting has been accounted for.
Eu\,II 4129\,{\AA} is the only europium line detected in the spectrum,
which consists of more than 30 components in total.
The Eu abundance of {\JrII} was thus derived by synthesizing the Eu\,II 4129\,{\AA} line,
taking the HFS data from \citet{Lawler2001ApJa} and the meteoritic isotopic abundance ratio
of $^{151}$Eu : $^{153}$Eu = 47.8 : 52.2.
As shown in Fig.~\ref{fig:abun_C2Eu}, abundance ratios of the second peak elements
relative to iron for {\JrII} are notably higher than normal EMP giants with similar metallicities.

\section{Heavy-element abundance pattern of {\JrII}}\label{sec:abun-pattern}

\begin{figure}
 \begin{center}
  \includegraphics[width=11cm]{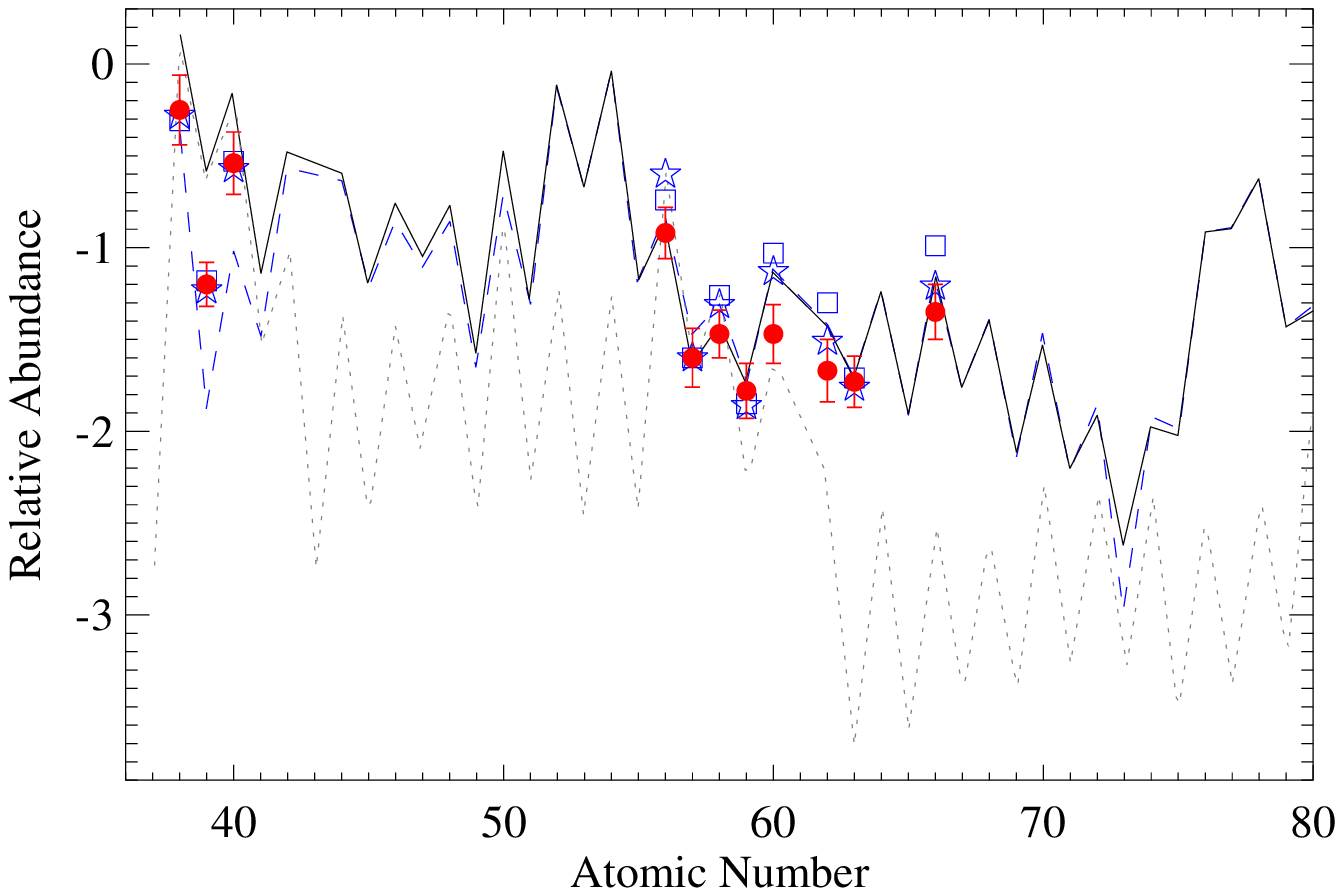}
 \end{center}
\caption{Heavy-element abundance patterns of {\JrII} (filled circles) and r-II stars
with {\Tefft} cooler than 5000\,K, including CS~22892$-$052 \citep[][squares]{Sneden2003ApJ},
and CS~31082$-$001 \citep[][stars]{Hill2002AA}.
The element abundances are scaled to match La in {\JrII}.
The solid curve refers to the SSr abundance patterns from \citet{Bisterzo2014ApJ},
and the dashed and dotted lines indicate those of the $r-$ and $s-$process components
in the SSr from \citet{Arlandini1999ApJ}, respectively.}\label{fig:abun_pattern}
\end{figure}

{\JrII} is the cool $r-$II giant with the lowest metallicity yet known.
Fig.~\ref{fig:abun_pattern} compares its abundance pattern of heavy elements
with those of two well-studied cool $r-$II giants with {\Tefft}$<$5000\,K.
CS~22892$-$052 \citep{Sneden2003ApJ}, and CS~31082$-$001 \citep{Hill2002AA}.
Abundances of the comparison stars are scaled to match the abundance of La in {\JrII}.
Previous studies have indicated that the pattern of the heavy-element abundances
in these two comparison stars are tracing the main component of the $r-$process \citep{Sneden2003ApJ,Roederer2014MNRAS}.
It is clear from the comparison that in the range from Sr to Dy,
the abundance pattern of the neutron-capture elements in {\JrII}
is very similar to the two cool $r-$II giants, and thus the main $r-$process.
For example, the dispersion of the average difference of log\,$\epsilon$(x) values
between {\JrII} and the two comprison $r-$II stars are about 0.19\,dex (vs. CS~22892$-$052)
and 0.15\,dex (vs. CS~31082$-$001), respectively,
which is comparable to the 1$\sigma$ error in our elemental abundance determinations.
The similar chemical abundance pattern observed in neutron-capture elements
suggest that there shall be a common origin of these elements in the classical $r-$process.

For comparison, the SSr patterns predicted by \citet{Bisterzo2014ApJ} and \citet{Arlandini1999ApJ}
are also displayed in Fig.~\ref{fig:abun_pattern}, respectively in solid and dashed lines.
These two sets of solar $r-$components are very much similar except elements
with significant contribution of $s-$process to their solar abundances,
especially for the light trans-iron elements such as Sr, Y, and Zr.
However, as discussed in \citet{Mashonkina&Christlieb2014AA}, it would be difficult to
firmly conclude on the relation of light trans-iron elements
between $r-$II stars and the solar $r-$process,
due to the large uncertainty in the solar $r-$residuals.
When compared to the predicted abundances by \citet{Bisterzo2014ApJ}
and those by \citet{Arlandini1999ApJ},
the elements in the range from Ba through Dy in {\JrII} are found to match
the scaled SSr pattern very well, with a dispersion of 0.13\,dex and 0.16\,dex
about the average abundance differences of the eight second peak elements, respectively.
The observed match is in line with previous studies on other $r-$process rich stars,
and provides additional evidence of universal production ratio
of these elements during the evolution of the Galaxy.

\begin{figure}
 \begin{center}
  \includegraphics[width=9cm]{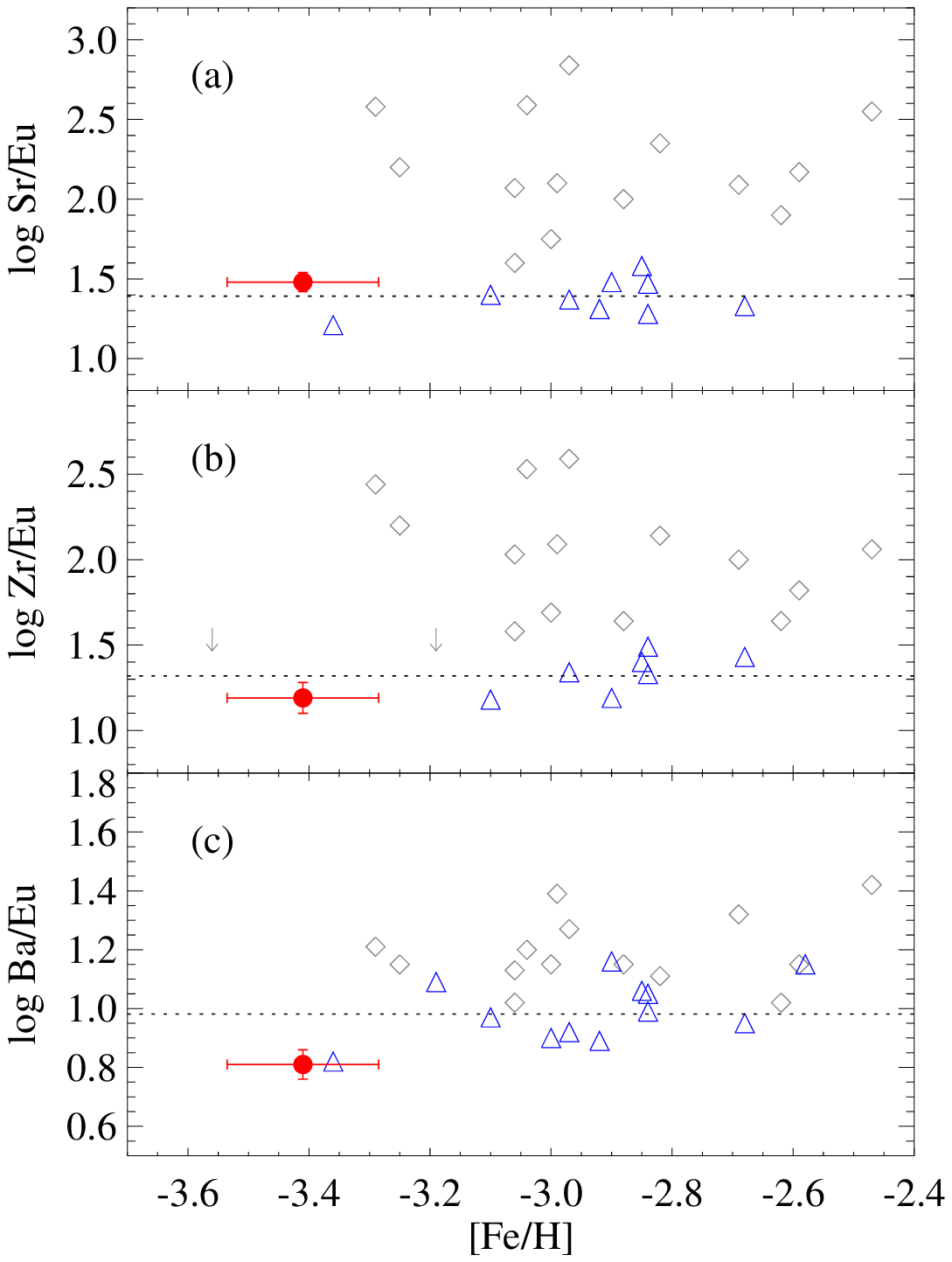}
 \end{center}
\caption{The distribution of the log\,Sr/Eu (a), log\,Zr/Eu (b), and log\,Ba/Eu (c)
abundance ratios in $r-$II stars ({\JrII} in filled circles, and literature $r-$II stars in triangles),
and regular EMP giants from ``First Stars''
\citep[][diamonds for abundances and arrows for upper limits]{Francois2007AA}.
The dotted lines indicate the average abundance ratio of the known $r-$II stars,
as described in the text.}\label{fig:abun_xEu}
\end{figure}

Previous studies on metal-poor halo stars suggest that there exist
distinct mechanisms between the production of light trans-iron elements
and that of heavier elements beyond Ba \citep{Aoki2005ApJ,Francois2007AA,Mashonkina2007ARep}.
By choosing Sr (Zr) and Eu to represent the first and second neutron-capture element peaks,
we could inspect the Sr/Eu and Zr/Eu abundance ratios of the $r-$II stars.
Fig.~\ref{fig:abun_xEu}a and \ref{fig:abun_xEu}b show the distribution of
abundance ratios of log\,Sr/Eu and log\,Zr/Eu of known $r-$II stars including {\JrII}.
The average value of log\,Sr/Eu and log\,Zr/Eu are 1.39$\pm$0.11 and 1.32$\pm$0.12 (dotted lines),
which are consistent with the results of previous studies \citep[e.g.][]{Mashonkina2010AA},
and are relatively smaller than those of normal EMP stars.
Also, an average value of log\,Ba/Eu of 0.98$\pm$0.11 (the dotted line in Fig.~\ref{fig:abun_xEu}c)
is derived from the 12 $r-$II stars, which well agrees with the predicted value of 0.93
for the pure $r-$process production of heavy elements \citep{Arlandini1999ApJ},
which indicates that the environment from which these stars were formed
only contained small amount of $s-$nuclei.
The distribution of the abundance ratios as shown in Fig.~\ref{fig:abun_xEu}
also suggests a common origin among the first and second $r-$process
peak elements in strongly $r-$process enhanced stars.

\section{Conclusion}\label{sec:conclusion}

{\JrII} has been selected as a candidate of EMP star from LAMOST spectroscopic survey,
and followed-up obtaining a high-resolution spectrum with Subaru/HDS.
The relatively high quality Subaru/HDS spectrum enables us to determine accurate parameters
and elemental abundances for 23 species, including 11 elements in the nuclear charge range of $Z=38-66$
which covers the light trans-iron and the second $r-$process peak elements.
Detailed abundance analysis confirms that {\JrII} is a strongly $r-$process enhanced EMP star
having \ABeq{Eu}{Fe}{+1.16}. It is also found that in the range from Sr through Dy,
{\JrII} presents very similar abundance patterns of the elements to
the well studied cool $r-$II giants, CS~22892$-$052 and CS~31082$-$001,
whose patterns of heavy elements can be well explained by a main component of the $r-$process.
Therefore {\JrII} is a newly discovered member of the small sample of currently known $r-$II stars,
with the lowest metallicity of \FeHsim{-3.4} among the $r-$II giants.

The abundance pattern from Ba through Dy in {\JrII} can be well matched
by the scaled Solar $r-$process pattern prediction by \citet{Arlandini1999ApJ} and \citet{Bisterzo2014ApJ}.
However, the large uncertainty of the Solar $r-$residuals
for the first $r-$process peak elements leads to difficulties to draw any conclusion
concerning the relation between the light trans-iron elements in {\JrII} (and other $r-$II stars)
and the Solar $r-$process.

Apart from different metallicities, {\JrII} presents quite similar abundance ratios
of Sr/Eu and Zr/Eu compared with previously well studied $r-$II stars,
and further confirms that {\JrII} belongs to the group of $r-$II stars.
{\JrII} turns to be the lowest metallicity $r-$II star with measured Zr abundance.
The extreme enhancement in \AB{Eu}{Fe} and low Sr/Eu (Zr/Eu, Ba/Eu as well)
in {\JrII} stars suggest a single or very few nucleosynthesis events
as is the case for other $r-$II stars.

However, due to limited wavelength coverage of the snapshot spectrum,
our analysis does not include any elements heavier than Dy.
To investigate abundance of the third $r-$process peak elements of {\JrII},
we will further obtain spectra with higher quality and higher resolution
which covers the UV band to investigate the chemical abundances of
heavier elements beyond Dy.

\normalem
\begin{acknowledgements}
We are grateful to the anonymous referee for helping improve the paper.
H.N.L. and G.Z. acknowledge supports by NSFC grants No. 11103030, 11233004, and 11390371.
W.A. and T.S. are supported by the JSPS Grant-in-Aid for Scientific Research (S:23224004).
S.H. is supported by the JSPS Grant-in-Aid for Scientific Research (c:26400231).
N.C. acknowledges support from Sonderforschungsbereich 881
``The Milky Way System'' (subproject A4) of the German Research Foundation (DFG).
Guoshoujing Telescope (the Large Sky Area Multi-Object Fiber Spectroscopic Telescope, LAMOST)
is a National Major Scientific Project built by the Chinese Academy of Sciences.
Funding for the project has been provided by the National Development and Reform Commission.
LAMOST is operated and managed by the National Astronomical Observatories, Chinese Academy of Sciences.
This work is based on data collected at the Subaru Telescope,
which is operated by the National Astronomical Observatory of Japan.
\end{acknowledgements}

\bibliographystyle{raa}
\bibliography{mpstar,lamost,rIIstar}

\end{document}